\documentclass[reprint,amsmath,amssymb,aps,prd,longbibliography,superscriptaddress,preprintnumbers]{revtex4-2}
\pdfoutput=1 
\usepackage{graphicx} 
\usepackage{bm} 
\usepackage{xcolor}
\usepackage[utf8]{inputenc}
\usepackage{amsmath, amsthm, amsfonts, amssymb}
\usepackage{braket}
\usepackage{tensor}
\usepackage{dsfont}
\usepackage{cancel}
\usepackage{mathrsfs}  
\usepackage{amsthm}

\begin{document}
 \title{Covariant non-perturbative pointer variables for quantum fields}
 
 \author{Alejandro Blanco S\'anchez}\affiliation{Departamento de F\'{\i}sica Te\'orica and IPARCOS, Universidad Complutense de Madrid, 28040 Madrid, España}
  \author{Luis J. Garay} 
 \affiliation{Departamento de F\'{\i}sica Te\'orica and IPARCOS, Universidad Complutense de Madrid, 28040 Madrid, España}
 
\author{Jose de Ram\'on} 
 \affiliation{\mbox{Departamento de Física, Universidad de Burgos, Pza. Misael Ba\~nuelos s.n., 09001 Burgos, España}}
  
  \begin{abstract}
We  describe the dynamics of a detector modeled by a harmonic oscillator  coupled with an otherwise  free quantum field in a curved spacetime in terms of covariant equations of motion leading to local observables. To achieve this, we derive and renormalize the integro-differential equation that governs the detector pointer-variable dynamics, introducing phenomenological parameters such as a dispersion coefficient and a Lamb-shift parameter. Our formal solution, expressed in terms of Green's functions, allows for the covariant, and causal analysis of induced observables on the field. This formalism can be used for instance to detect non-Gaussianities present in the field's state.

\end{abstract}

\date{February 3, 2025}
\preprint{IPARCOS-UCM-25-009}
\maketitle

\tableofcontents
 
\section{Introduction}\label{sec: intro}
 
Quantum field theory (QFT) \cite{Schwartz:2014sze}  accounts for relativistic quantum phenomena. One of the main applications of QFT is in high energy physics, where one is most often concerned with measuring $S$-matrix elements defined at asymptotic spacetime regions under the assumption that the involved parties do not interact in these limits \cite{Blum2017-BLUTSI}. For this reason, at least in the developemental stages of the theory, not much attention was paid to formalizing a local measurement theory for QFT.

We now know  that there are certain instances where we would be interested in performing local rather than asymptotic measurements, for instance when measuring Bell-type correlations \cite{papageorgiou2023eliminating,PhysicsPhysiqueFizika.1.195,Summers:1987fn}. To this avail, particle detectors are defined as purely quantum mechanical objects that aid in formalizing this local measurement theory and provide insight into the structure of quantum fields \cite{fewster2020quantum,PhysRevD.105.065003}.
They allow us to probe the structure of quantum fields and obtain direct information about them by measuring the detector, which is a simpler system and easier to prepare practically. Also, given that the concept of particle in quantum field theory is not well defined \cite{Wald:1984rg}, detectors allow us to define a particle as what a particle detector measures \cite{Pipo}.

One of the first particle detector proposed was the Unruh-DeWitt model \cite{Unruh:1976db,DeWitts} where a two-level system is used to probe the structure of a quantum field. This kind of detector models has played a role in the analysis of several phenomena in QFT, such as the Unruh effect where it can be proven that such detectors thermalize with the field in a (3+1)-dimensional spacetime when the detector follows a uniformly accelerated trajectory, with a temperature proportional to the acceleration \cite{Fewster_2016,Unruh:1976db}.

Historically, due to its simplicity and its success in describing several phenomena, the analysis of particle detectors is commonly restricted to the Unruh-DeWitt model, using a two-level quantum system for the detector and treating it perturbatively, but there is no reason to default to this option.

Recent works (see e.g.\cite{Pipo,PhysRevD.87.084062,Hu_2012,Torres2023}) have considered quantum harmonic oscillators as viable alternatives to the Unruh-DeWitt detector.
An advantage of using a quantum harmonic oscillator for our detector is that it allows for a more complex probing of the field. Indeed its associated Hilbert space is infinite dimensional and as such allows for the definition of what will be called a pointer variable that will give more precise information of the field than the simple ``click" of the Unruh-DeWitt detector. However, the main advantage of using a quantum harmonic oscillator as a detector is that it can be treated, at least formally, in a completely nonperturbative approach when working in the Heisenberg picture~\cite{CALDEIRA1983587}.

The fact that the natural dynamics of the oscillator is nonrelativistic often jeopardizes the ability of these models to make fully relativistic predictions, in particular in situations in which relativistic causality is relevant \cite{PhysRevD.103.085002}. It can be shown that these frictions with causality are related to the dynamics of the field within the spatial extension of the detector, and therefore one concludes that in order to make relativistic measurements with nonrelativistic systems one has to work in the approximation in which they are arbitrarily small, i.e. pointlike.  However, coupling a pointlike system to a quantum field is not devoid of problems. Indeed, when studying the coupling of our particle detector to the quantum field, one finds pathologies in the equations governing the dynamics of the pointer variable in the limit where the spatial width of the detector is zero.  Similar issues are well known, e.g., when considering classical pointlike particles interacting with electromagnetic fields \cite{Schwartz:2014sze}.   Mathematically, one can describe this issue in the context of distribution theory. As we will see, the dynamics of pointlike models require the evaluation of a specific distribution on the detector's trajectory, and the action of this distribution will not be well defined on every test function. To get rid of this divergence we will use a procedure known as Epstein-Glaser renormalization \cite{AIHPA_1973__19_3_211_0,Prange:1997iy} to characterize the extensions of the ill-defined distribution, thus leading to an equation of motion for the pointer variable that will depend on some phenomenological constants but that will be well defined.  The renormalized model, rather than a correction, should be thought of as an entirely independent model which cannot be derived directly as an approximation.

Once we have performed this renormalization, we will attempt to answer the question of what properties of the field are really being measured when measuring the pointer variable by means of an {induced observable} by the detector on the field, a concept introduced in \cite{busch}, and that has been applied to quantum field theoretical systems in \cite{fewster2020quantum}.  In short,   induced observables by the detector on the field are the field observables that explain the expectation value of an arbitrary detector observable after the interaction between them takes place. 

Hence, this work aims to build a simple framework for describing non-perturbative local measurements. There are specific reasons for this, beyond the general motivation. To mention two concrete problems, perturbative methods cannot be used to describe measurements violating Bell inequalities \cite{Summers:1987fn}; they cannot be used to distinguish non-Gaussian states (most remarkably for pure states) either. Therefore, a tractable formalism to describe these phenomena, which are concrete instances of deviations from classicality, while retaining locality and causality, is still missing. 

{
This work is organized as follows. In Section~\ref{sec: model} we introduce the detector model and state the assumptions for the joint dynamics of detector and field. In Section~\ref{sec: pvd} we derive an equation that describes non-perturbatively the evolution of the pointer variable in its proper time. We will see that the resulting integro-differential equation requires a renormalization procedure, which we will fully characterize. 
 In Section~\ref{sec: indu} we introduce the concept of induced observable by the detector on the field and particularize it to our model.
Section~\ref{sec:nomem} is devoted to the cases in which the pointer-variable satisfies a differential equation, as happens for instance for massless fields in Minkowski spacetime. Finally, in Section~\ref{sec: ng} we will provide a method for measuring non-Gaussianities of the field using a detector. We summarize and conclude in Section~\ref{Sec: Conclusion}.
}

\textbf{Notation:} We will use the signature $(-,+,+,+)$, units such that $\hbar=c=1$. We will denote spacetime event as $\mathsf{x}$ and spatial vectors as $\mathbf{x}$, and use the following convention for the Fourier transform: \mbox{$\tilde{f}[\omega] = (2\pi)^{-1/2}\int_\mathbb{R} dt f(t)e^{-i\omega t}$}. We also define the convolution of two functions $f$ and $g$ as \mbox{$
    f*g(x) = \int_{\mathbb{R}} dy f(y-x)g(y)$}.

\section{The model}\label{sec: model}
The goal of this section is to present the equations governing the dynamics of a real scalar quantum field $\hat{\Phi}$ coupled to a harmonic oscillator described by the pointer variable $\hat{Q}$ taken to be one of its quadratures. We will refer to the harmonic oscillator as a detector as it will be used to probe the structure of the field.

We will work within the Heisenberg picture and restrict our analysis to a (3+1)-dimensional globally hyperbolic spacetime $\mathcal{M}$, characterized by a metric tensor $\mathsf{g}$. This means that there exists a global time function $\tau(\mathsf{x})$ and the spacetime admits a foliation in Cauchy hypersurfaces $\Sigma_\tau$ defined by constant $\tau$. We will take the global time function to coincide with the detector's proper time along its trajectory $\mathsf{z}(\tau)$ and consider spatial vectors as vectors lying entirely within the Cauchy hypersurfaces, that is, $\mathbf{x} \in \Sigma_\tau$.

We will assume that the pointer variable $\hat{Q}(\tau)$ and the field are uncorrelated before a certain starting time, i.e., before the interaction between the field and the detector takes place, their state can be described as a product
\begin{equation}
    \hat{\rho}_0 = \hat{\rho}_D\otimes\hat{\rho}_\phi,\label{eqn: inicial}
\end{equation}
where $\hat{\rho}_D$ represents the detector's state and $\hat{\rho}_\phi$ that of the field. The field will be acting on a Hilbert space taken to be a symmetric Fock space \cite{Birrell_Davies_1982}, $\mathcal{H}_\phi$, and the detector will be acting on the usual harmonic oscillator's Hilbert space \cite{Cohen-Tannoudji:101367}. We will consider that the interaction takes place in a finite time determined by a smooth switching function with compact support $\chi(\tau)$, i.e. $\chi\in \mathcal{D}(\mathbb{R})$ is a test function.

The dynamics of our system is encoded in a classical action \cite{Wald:1984rg} for the field $\Phi$ and the harmonic oscillator ${Q}(\tau)$, and an interaction term $S_{\text{int}}$ between them
\begin{align}
    S=&-\frac{1}{2} \int_\mathcal{M} dV\left((\nabla \Phi)^2+m^2 \Phi^2\right) \nonumber\\
    &+ \frac{1}{2}\int_{\mathbb{R}}d\tau(\dot{{Q}}^2 - \Omega_0^2{Q}^2) + S_{\text{int}} ,
\end{align}
where  $m$ represents the  mass of the field, $\Omega_0$ is the characteristic frequency of the detector, $dV=\mathrm{d}^4 x \sqrt{-\mathfrak{g}}$ is the spacetime volume element, and $\mathfrak{g}$ is the determinant of the metric tensor $\mathsf{g}$. 

We will consider an interaction term of the form
\begin{equation}
    S_{\text{int}} = -\lambda \int_{\mathbb{R}} d\tau\chi(\tau){Q}(\tau){\Phi} (\mathsf z(\tau)),
\end{equation}
where $\lambda$ is a dimensionless coupling constant, $\chi$ is the switching function, and ${\Phi}(\mathsf{z}(\tau))$ is the pull-back of the field to the detector's trajectory $\mathsf{z}(\tau)$. 

Even though we are considering a classical action to describe the dynamics of our system, as it is quadratic in the canonical variables, the equations of motion derived from it are equivalent to the quantum equations obtained from Heisenberg's equations of motion. This allows for a fully classical treatment of the dynamics until we need to specify an initial quantum state. Nevertheless, we will choose to treat the system as quantum mechanical from here on to make the notation more compact. By performing a classical treatment and then promoting the canonical variables to operators acting on the full product Hilbert space, the dynamics of the joint system are given by the following equations:
\begin{align}
    \ddot{\hat{Q}}(\tau) + \Omega_0^2\hat{Q}(\tau) =& -\lambda\chi(\tau)\hat{\Phi}( \mathsf{z}(\tau)) ,\label{eqn: quadrature}\\
    (\square + m^2)\hat{\Phi}(\mathsf{x}) =& -\lambda\int_{\mathbb{R}} ds \frac{\chi(s)}{\sqrt{-\mathfrak{g}}}\delta^{(4)}(\mathsf{x} - \mathsf{z}(s))\hat{Q}(s).\label{eqn: field}
\end{align}
These two equations are clearly fully covariant as we are not making any reference to a specific coordinate system. Note that  $\hat Q$ and $\hat \Phi$ are operators that act on the whole Hilbert space.

\section{Pointer variable dynamics}\label{sec: pvd}
In this section we obtain a formal solution to equations \eqref{eqn: quadrature} and \eqref{eqn: field} for the pointer variable $\hat{Q}(\tau)$. These equations are ill-defined even in the distributional sense and will require renormalization. Particularly, we will employ a technique known as Epstein-Glaser renormalization \cite{AIHPA_1973__19_3_211_0}.

\subsection{Bare dynamics of the pointer variable}

Equation \eqref{eqn: field} is linear and thus admits a formal solution in terms of a Green function. To preserve causality, we take this Green function $G(\mathsf{x};\mathsf{y})$ to be the retarded propagator for the Klein-Gordon theory, i.e., it satisfies
\begin{equation}
    (\square_\mathsf{y} + m^2)G(\mathsf{x};\mathsf{y}) = -\frac{\delta^{4}(\mathsf{x} - \mathsf{y})}{\sqrt{-\mathfrak{g}}}
\end{equation}
and has support in the causal future of $\mathsf{x}$.
Then we may present a formal solution for $\hat{\phi}$ in the coordinate basis $(\tau,\mathbf{x})$ as
\begin{equation}
    \hat{\Phi}(\tau,\mathbf{x}) = \hat{\Phi}_0(\tau,\mathbf{x}) - \lambda\!\int_{\mathbb{R}} ds G(\tau, \mathbf{x}; s, \mathbf{z}(s))\chi(s)\hat{Q}(s),\label{eqn: campo}
\end{equation}
where $\hat{\Phi}_0$ is a solution to the free (homogeneous) Klein Gordon equation and whose precise form for an arbitrary globally hyperbolic spacetime can be found in \cite{Fulling_1989}.

Substituting \eqref{eqn: campo} into \eqref{eqn: quadrature} we find the equation for the pointer variable  
\begin{equation}
    \ddot{\hat{Q}} + \Omega_0^2\hat{Q} - {\lambda^2}\chi(\tau)\!\int_{\mathbb{R}} ds G(\tau, s)\chi(s)\hat{Q}(s)  = -\lambda\chi(\tau)\hat{\phi}_0(\tau),\label{eqn: norenormalizada}
\end{equation}
where we have defined $G(\tau,s) = G(\tau, \mathbf{z}(\tau); s,\mathbf{z}(s))$ as the pull-back of the retarded propagator to the detector's trajectory and likewise for $\hat\phi_0(\tau)=\hat{\Phi}_0(\mathsf{z}(\tau))$.
The source is the pull-back of the free field operator to the detector's trajectory modulated by the switching function. This equation is not well defined in general because, in a distributional sense, the retarded propagator cannot be evaluated at the trajectory of the detector, since it diverges in the coincidence limit $\tau = s$.

\subsection{Need for renormalization}

We will not formally prove that this distribution is generally ill defined, but we will use the example of a  massless field in Minkowski spacetime to illustrate where and why issues arise.
In this case the retarded propagator takes the functional form \cite{Fulling_1989}
\begin{equation}
    G(\mathsf{x},\mathsf{y}) = \frac{1}{4\pi}\theta(\mathsf{x}^0 - \mathsf{y}^0)\delta[(\mathsf{x}-\mathsf{y})^2].\label{eqn: retrasado}
\end{equation}

If we consider an inertial trajectory for the detector $\mathsf{z}(t) = (t,\mathbf{0})$, we can see that the propagator is ill defined through its action on test functions. To see this, we will consider a special type of test function   of the form 
\begin{equation}
   f(\mathsf{x}) = F_L(\mathbf{x})g(t),
\end{equation}
where $g(t)$ is a one-dimensional test function, $F_L$ is a function such that, for any positive $L$,
\begin{equation}
    F_L(\mathbf{x}) = \frac{1}{L^3}F_1(\mathbf{x}/L),
\end{equation}
and $F_1$ is a 3-dimensional test function of dimensionless argument with $\int_{\mathbb{R}^3}d^3\mathbf{x}F_1(\mathbf{x}) = 1$.
Note that in the limit   $L\rightarrow 0$, $F_L(\mathbf{x})\rightarrow\delta^{3}(\mathbf{x})$ in the distributional sense and therefore the action of the propagator in \eqref{eqn: retrasado} on $f$ will act as the pull-back to an inertial trajectory of constant~$\mathbf{x}$.

The retarded propagator at the detector's inertial trajectory   acts upon these test functions as 
\begin{align}
    G[f,f]
    = \int_\mathcal{M}\!\!\!d^4\mathsf{x}\int_\mathcal{M} \!\!\!d^4\mathsf{x}^\prime G[t,\mathbf{x};t^\prime,\mathbf{x}]  F_L(\mathbf{x})F_L(\mathbf{x})g(t)g(t^\prime).
\end{align}
Taking into account that the retarded propagator is translationally invariant, this equation can be written in terms of convolutions as
\begin{align}
    G[f,f]
    = \int_\mathcal{M}\!\!\!dtd^3\mathbf{x} \;G(t,\mathbf{x})  \;\tilde{F}_L*F_L(\mathbf{x})\;\tilde{g}*g(t),\label{eqn: kla}
\end{align}
where $G(\mathsf{x}) = G(\mathsf{x};0)$ and the $*$-operation represents the convolution. From the properties of the $\delta$-function it follows that the retarded propagator \eqref{eqn: retrasado} can also be written as 
\begin{align}
    G(\mathsf{x}) = \frac{1}{8\pi t}\delta(t - \|{\mathbf{x}}\|),
\end{align}
which means that we can perform the spatial integral in~\eqref{eqn: kla} in spherical coordinates. This leads to
\begin{align}
    G[f,f] = \frac{1}{2}\int_0^\infty dt t\;\tilde{g}*g(t)\; [ \tilde{F}_L*F_L]_{\text{av}}(t),
\end{align}
where we have defined the spherical average $[\cdot]_{\text{av}}$ as the integral in the angular variables and we have evaluated the $\delta$-function. It is straightforward to see that 
\begin{align}
    [ \tilde{F}_L*F_L]_{\text{av}}(t) = \frac{1}{L^3}[ \tilde{F}_1*F_1]_{\text{av}}(t/L).
\end{align}
Performing the change of variable $t \rightarrow u = t/L$, it follows that
\begin{align}
    G[f,f] = \frac{1}{2L}\int_0^\infty du u\;\tilde{g}*g(Lu)\; [ \tilde{F}_1*F_1]_{\text{av}}(u).
\end{align}
It is now  clear that this expression is divergent in the limit where the arbitrary scale $L\rightarrow 0$, since the expression under the integral sign is finite in this limit.

Therefore, in this particular case we have found that the pull-back of the retarded propagator to the detector's trajectory is not in general a well-defined distribution. From \eqref{eqn: retrasado} it is clear that the issues arise in the coincidence limit $\mathsf{x} = \mathsf{y}$. Actually it can be shown that this issue persists for the pull-back of propagators of Klein-Gordon fields to general trajectories in general globally hyperbolic spacetimes \cite{Takagi:1986kn}. In the limit $\tau \to s$, it can be seen that for a general (3+1)-dimensional spacetime \cite{Fulling_1989} the retarded propagator is translationally invariant and has the same divergent structure as the also ill-defined distribution
\begin{equation}
    \mathcal{T}(\tau,s) =\theta(\tau - s)\delta[(\tau - s)^2].\label{eqn: divergenciaretard}
\end{equation}

This allows us to formally define the integral kernel
\begin{equation}
    K(\tau, s) = G(\tau,s) - \frac{1}{4\pi}\mathcal{T}(\tau,s).
\end{equation}
Even though this object is only defined formally (it is the difference of two ill-defined distributions), it can be shown \cite{Fulling_1989} that the action of $K(\tau,s)$ on arbitrary test functions is actually well defined as we have subtracted the divergent part of the retarded propagator, with $K(\tau,s)$ having at most a logarithmic divergence. The effect of this logarithmic divergence in \eqref{eqn: norenormalizada} disappears upon integration. 
The exact form of this integral kernel depends on the spacetime under consideration and the field's mass \cite{Fulling_1989}. For example, for a massless scalar field in   (3+1)-dimensional Minkowski  spacetime, $K(\tau,s) = 0$.

Given that the retarded propagator is not a well-defined distribution, the bare integro-differential equation \eqref{eqn: norenormalizada} is not well defined and a renormalization procedure is required. However, based on the observations made above, we can 
write the retarded propagator $G$ as the sum of a finite part $K$ plus the divergent one $\mathcal T$. We then proceed to renormalize $\mathcal T$ and
define a renormalized propagator independent on the particulars of the trajectory and the globally hyperbolic spacetime by writing    
\begin{align}
    G_{\textsc{r}}(\tau,s) = K(\tau,s) + \frac{1}{4\pi}\mathcal{T}_{\textsc{r}}(\tau,s),
\end{align}
where the sub-index R denotes the renormalized distribution.

\subsection{Epstein-Glaser renormalization}

As mentioned above, the issue with the distribution $\mathcal{T}(\tau,s) $, whose action on test functions $f$ is given by
\begin{equation}
    \mathcal{T}^\tau[f] =\int_{\mathbb{R}}ds\mathcal{T}(\tau,s)f(s),
\end{equation}
arises in the coincidence limit $\tau = s$, that is, the divergence of the distribution is contained in a single real point for each $\tau$.
With this in mind, we will follow the renormalization procedure introduced by Epstein and Glaser~\cite{AIHPA_1973__19_3_211_0}, which was developed to address this type of divergences. We will follow the notation and the derivations described in Ref.~\cite{Prange:1997iy}. 

Let   $\mathcal{D}^\prime(\mathbb{R})$ be the space of distributions acting on test functions $f \in\mathcal{D}(\mathbb{R})$. Formally, our goal is to construct a distribution $\mathcal{T}^\tau_\textsc{r}$ such that for every $f \in \mathcal{D}(\mathbb{R}\setminus\{\tau\})$,
\begin{equation}
    \mathcal{T}^\tau[f] = \mathcal{T}^\tau_\textsc{r}[f],
\end{equation}
and characterize its possible extensions to $f \in \mathcal{D}(\mathbb{R})$. Let us take $\tau = 0$, for the time being.

It can be shown \cite{Prange:1997iy} that a given distribution $\mathcal{T}^0$ in \mbox{$\mathcal{D}^\prime(\mathbb{R}\setminus\{0\})$} can only have  finitely many extensions to $\mathcal{D}^\prime(\mathbb{R})$ with 
the same scaling degree at $0$, defined as the infimum of all  $\nu$ such that 
\begin{equation}\lambda^{\nu}\mathcal{T}^0(\lambda s) \overset{\lambda\rightarrow 0}{\longrightarrow} 0  ,
\end{equation}
where the limit is taken in the sense of distributions.

In particular, all the extensions are characterized by 
\begin{align}
    \mathcal{T}^0_\textsc{r}[f] = \mathcal{T}^0[ W [f]] + \sum_{k = 0}^{[\nu-1]} C_k\delta^{(k)}[f],
\end{align}
where $\delta^{(k)}$ are distributional derivatives of the $\delta$-function, $C_k$ are a set of arbitrary constants, and $[x]$ represents the largest natural number smaller than $x$. $W$ is defined as  a smooth subtraction to the test function $f$ in the singular point of the original distribution, as follows:
\begin{equation}
    W[f](s)=f(s)-w(s) \sum_{k=0} ^{[\nu-1]}
    \frac{\partial^k}{\partial s^k}
    \frac{f}{w}\bigg|_{s=0}\cdot\frac{s^k}{k!},
\end{equation}
with $w\in\mathcal{D}(\mathbb{R})$ and $w(0) \neq 0$. These subtracted test functions vanish as $s^{[\nu]}$ when $s\to 0$ .

In the (3+1)-dimensional case, the divergent part of the propagator satisfies for $\lambda\rightarrow 0$,
\begin{align}
    &\lambda^{\nu} \int_{\mathbb{R}} \!ds\theta(-\lambda s)\delta(\lambda^2 s^2)f(s)\nonumber\\  &=  \lambda^{\nu-2}\int_{\mathbb{R}}\!ds\theta(-s)\delta( s^2)f(s) \rightarrow 0 ,
\end{align}
for $\nu \geq 2$,
so the scaling degree of $\mathcal{T}^0$ at $s = 0$ is $\nu = 2$.
Therefore, its possible renormalizations with this scaling degree are  
\begin{equation}
    \mathcal{T}_\textsc{r}^0[f]=  \mathcal{T}^0[ W [f]] + C_0f(0) + C_1\dot{f}(0),\label{eqn: hj}
\end{equation}
where $C_0$ and $C_1$ are arbitrary constants. The value of these constants cannot be extracted from the bare model, and have to be fixed {phenomenologically}, as usually happens when renormalizing a field theory. Furthermore, 
given that the divergent part of our distribution only has support in the origin, $
    \mathcal{T}^0[W[f]] = 0$ and, therefore,
\begin{equation}
    \mathcal{T}_\textsc{r}^0[f]=  C_0f(0) + C_1\dot{f}(0),
\end{equation}
so that
\begin{equation}
    \mathcal{T}_\textsc{r}(0,s) = C_0\delta(s) + C_1\dot{\delta}(s).
\end{equation}
In other dimensions the universal divergent part of the propagator need not be zero for timelike vectors and therefore, the renormalization procedure may be more involved.

We did this calculation taking $\tau = 0$ but every idea still applies for an arbitrary $\tau$ as the divergent structure of $\mathcal{T}^\tau$ in $\tau$ is the same as that of $\mathcal{T}^0$ at $0$. Therefore, for any $\tau$, $\mathcal{T}_\textsc{r}(\tau,s)$ takes the form 
\begin{equation}
    \mathcal{T}_\textsc{r}(\tau, s) = C_0(\tau)\delta(\tau - s) + C_1(\tau)\dot{\delta}(\tau - s),
\end{equation}
where $C_0$ and $C_1$ are two arbitrary functions of $\tau$. However, for the purposes of this work, we will assume that these functions are independent of the time $\tau$ in which we perform the renormalization. This assumption is based on the observation that the ultraviolet contributions to the dynamics should  essentially be the ones for inertial detectors in flat spacetime, according to   general-covariance  and that such interaction should be translationally invariant in time. On the other hand, it could also be argued, given the singularity of the interaction, that these constants could have some time dependence through some invariants associated with the trajectory, such as the acceleration and the curvature \cite{Brunetti2000}. If this were  the case, it would not affect the results of this paper but simply add an extra time-dependence in the time-dependent functions $\Gamma(\tau)$ and $\Omega(\tau)$ defined below.

Under the assumption of time independence, we define
$C_0=\Delta\Omega_0^2$ and $\Gamma_0=C_1/2$ and write
\begin{equation}
    \mathcal{T}_\textsc{r}(\tau, s) = \Delta\Omega_0^2\delta(\tau - s) + 2\Gamma_0\dot{\delta}(\tau - s).\label{eqn: renormalizado}
\end{equation}
The reasoning behind this redefinition of the constants will become clear in the following.

\subsection{Renormalized dynamics}

 After renormalizing the retarded propagator, one can define the renormalized dynamics by simply    substituting the integral kernel in \eqref{eqn: norenormalizada} by its renormalized version. This yields 
\begin{align}
  &   \ddot{\hat{Q}}(\tau)  + 2\Gamma(\tau)\dot{\hat{Q}}(\tau) + \Omega^2(\tau)\hat{Q}(\tau)\nonumber\\
    & - {\lambda^2}\int_{\mathbb{R}} ds K(\tau,s)\chi(\tau)\chi(s)\hat{Q}(s)  =-{\lambda}\chi(\tau)\hat{\phi}_0(\tau),\label{eqn: finalintegrodif}
\end{align}
where
\begin{align}\label{gammalamb}
    \Gamma(\tau)& =  \lambda^2\Gamma_0\chi^2(\tau),\\
    \Omega^2(\tau)&  =  \Omega_0^2 + \lambda^2\Delta\Omega_0^2\chi^2(\tau) + \dot{\Gamma}(\lambda,\tau).
\end{align}

The integral kernel $K(\tau,s)$ is a regular and integrable function that depends on the spacetime under consideration, and the mass of the field. As mentioned above, the arbitrary constants that come from the renormalization procedure are phenomenological constants that are associated with the damping and the frequency shift of the pointer variable. Note that in the case $\chi = 1$, $\Gamma_0$ coincides exactly with the damping coefficient of a damped harmonic oscillator and $\Delta\Omega_0$ can be interpreted as a Lamb shift.

\section{Induced observables}\label{sec: indu}

In this section we will introduce the concept of induced observable by the detector on the field for a general detector model and we will particularize this definition to our specific model. This analysis can be also found in~\cite{fewster2020quantum}, in which they also discuss   couplings which are linear in both otherwise free systems, and the induced observables in such scenarios. Our results depart from those in \cite{fewster2020quantum}  in the following sense: our induced observables can be written in terms of more tractable distributions that can be computed, e.g., numerically.

\subsection{General detector state}

The notion of induced observable comes from the assumption that the observables that are being measured act only on the detector's Hilbert space. When these observables evolve in the Heisenberg picture, however, they will act on the full product Hilbert space, since the field and the detector are interacting. The induced observables are simply the observables of the field that are being measured given measurements of the detector. 

The observables associated with the detector before the interaction starts are always of the form 
\begin{equation}
    \hat{A}_\text{D}(0) = \hat{A}_\text{D}\otimes \mathds{1}_\phi,
\end{equation}
acting trivially on the field's Hilbert space. After the interaction they will act on the full Hilbert space and take the form $\hat{A}_{\text{D}}(\tau) = \hat{U}^\dagger(\tau) (\hat{A}_{\text{D}}\otimes\mathds{1}_{\text{D}})\hat{U}(\tau)$, where $\hat{U}(\tau)$ is the joint evolution. 

The expectation value of these observables after the interaction, assuming that the initial state is uncorrelated, is given by
\begin{equation}
    \braket{\hat A_{\text{D}}(\tau)} = \text{Tr}[(\hat{\rho}_\text{D}\otimes\hat{\rho}_\phi) \hat{A}_\text{D}(\tau)].
\end{equation}
where this full trace can be taken as two partial traces over the detector's and the field's Hilbert spaces.

This means that one can define the following map $\mathcal{E}$ that takes linear operators $\hat{A}_\text{D}$ acting on the detector's Hilbert space to operators $ \mathcal{E}_\tau(\hat{A}_{\text{D}})$ acting on the Hilbert space  of the field:
\begin{equation}
    \mathcal{E}_\tau(\hat{A}_{\text{D}}) = \operatorname{Tr}_\text{D}[(\hat{\rho}_\text{D}\otimes\mathds{1}_{\phi})\hat{A}_{\text{D}}(\tau)].\label{eqn: induceddef}
\end{equation}
This field operator directly satisfies
\begin{equation}
    \braket{\hat{A}_{\text{D}}(\tau)} = \operatorname{Tr}_\phi(\hat{\rho}_\phi\mathcal{E}_\tau[\hat{A}_{\text{D}}]).
\end{equation}
and receives the name of {induced observable by the detector on the field}.

For the specific model we are considering in this work, we will calculate the observables induced by measurements of the pointer variable. To do so, it is useful to consider the characteristic function for the pointer variable, 
\begin{equation}
    M_{\hat{Q}}(\mu;\tau) = \braket{e^{i\mu\hat{Q}(\tau)}}.
\end{equation}
This quantity is in general, not easy to compute as we need a solution $\hat{Q}(\tau)$ of \eqref{eqn: finalintegrodif}, and there is no known procedure to obtain an analytic solution to this equation. Despite this, given that the equation for $\hat{Q}(\tau)$ is linear, we may construct a formal solution for $\hat{Q}(\tau)$ and capture some essential aspects of  $M_{\hat{Q}}(\mu;\tau)$ by means of fundamental solutions. 

Indeed, Eq.~\eqref{eqn: finalintegrodif} has a formal solution of the form
\begin{equation}
    \hat{Q}(\tau) = \hat{Q}_0(\tau) +\hat\varphi_0(\tau),
    \label{eq:generalsol}
\end{equation}
where the first term acts non-trivially only on the detector's Hilbert space and fulfills the homogeneous integro-differential equation and the second term 
\begin{equation}
\hat\varphi_0(\tau)=- \lambda R[\chi\hat{\phi}_0](\tau)=-\lambda \int_\mathbb{R} ds R(\tau,s)\chi(s)\hat\phi_0(s)
\end{equation}
acts non-trivially only on the field's Hilbert space, where $R$ is taken as the retarded Green function of the homogeneous equation associated with \eqref{eqn: finalintegrodif}. We may then write 
\begin{equation}
    M_{\hat{Q}}(\mu;\tau)  = \braket{e^{i\mu\hat{Q}_0(\tau) +i\mu\hat\varphi_0(\tau)}}.
\end{equation}
As we are taking the expectation value with respect to the initial product state $\hat{\rho}_\text{D}\otimes\hat{\rho}_\phi$, this expression can be further simplified to
\begin{equation}\label{characteristic}
    M_{\hat{Q}}(\mu;\tau) = \braket{e^{i\mu\hat{Q}_0(\tau)}}_\text{D} \braket{e^{ i\mu\hat\varphi_0(\tau)}}_\phi,
\end{equation}
where the first expectation value is taken with respect to $\hat{\rho}_\text{D}$ in the detector's Hilbert space and the second one with respect to $\hat{\rho}_\phi$ in the field's Hilbert space.

The reason why we are interested in the characteristic function is because we will consider observables of the pointer variable in the Schrödinger picture $\hat{Q}$ of the form
\begin{equation}
    \xi(\hat{Q})= \frac{1}{\sqrt{2\pi}}\!\int_{\mathbb{R}}d\mu \tilde{\xi}(\mu)e^{i\mu\hat{Q}},
\end{equation}
where $\xi$ is a certain density function and $\tilde{\xi}$ denotes its Fourier transform. In the Heisenberg picture, this Schrödinger observable evolves in time as 
\begin{equation}
    \xi(\hat{Q})(\tau) =  \frac{1}{\sqrt{2\pi}}\int_{\mathbb{R}}d\mu\tilde\xi(\mu)e^{i\mu\hat{Q}(\tau)},
\end{equation}
where $\hat{Q}(\tau)$ is the solution to Eq.~\eqref{eqn: finalintegrodif}. 
Taking the expectation value,
\begin{align}
    \braket{\xi(\hat{Q})(\tau)}   &= \frac{1}{\sqrt{2\pi}}\int_{\mathbb{R}}d\mu \tilde{\xi}(\mu)\braket{e^{i\mu\hat{Q}_0(\tau)}}_\text{D} \braket{e^{  i\mu\hat\varphi_0(\tau)}}_\phi.\label{eqn: expectacion}
\end{align}
Consequently, these observables induce the following observables on the field,
\begin{equation}
    \mathcal{E}_\tau[\xi(\hat{Q})]  = \frac{1}{\sqrt{2\pi}}\int_{\mathbb{R}}d\mu \tilde{\xi}(\mu)\braket{e^{i\mu\hat{Q}_0(\tau)}}_{\text{D}} e^{  i\mu\hat\varphi_0(\tau)}.\label{eqn: induced}
\end{equation}
Note that these induced observables of the field are completely determined by the homogeneous dynamics, the retarded Green function, and the characteristic function of the detector in the initial state.

\subsection{Detector initially in the  ground state }

Let us concentrate  in the solution $\hat{Q}_0$ of the homogeneous equation. The general solution is of the form 
\begin{equation}\label{QW}
\hat{Q}_0=p(\tau)\big(\hat{a} u(\tau)+\hat{a}^\dagger u^*(\tau)\big),
\end{equation}
where we have defined a damping factor
\begin{equation}
        p(\tau)=e^{-\int_{-\infty}^\tau d s\Gamma(s)}
\end{equation}
and   $u$  is a complex solution of the equation
\begin{equation}\label{lauk}
    \ddot u+\omega^2(\tau)u- \big({\lambda^2}\chi(\tau)/p(\tau)\big)\int_{\mathbb{R}} ds K(\tau,s)\chi(s)p(s) u(s) =0,
\end{equation}
 normalized so that $u(\tau)\dot u^*(\tau)-u^*(\tau)\dot u(\tau)=i$ when $\tau$ is a time before the interaction has started. If the kernel $K$ vanished, this normalization would hold for any $\tau$, but this is no longer true in the presence of the kernel $K$.
  We also have defined the time-dependent frequency
\begin{align}
    \omega(\tau)^2&=\Omega(\tau)^2-\dot \Gamma(\tau)-\Gamma(\tau)^2
    \nonumber\\
    &= \Omega_0^2 + \lambda^2(\Delta\Omega_0^2-\Gamma^2_0)\chi^2(\tau) .
\end{align}
The character of Eq.~\eqref{lauk} depends strongly on the phenomenological parameters and the switching function~$\chi$.  The kernel-free part of the equation will describe a harmonic oscillator with time-dependent frequency $\omega^2>0$ if
\begin{align}\label{ineqk}
    \Gamma_0^2<\frac{\Omega_0^2}{\chi^2(\tau)\lambda^2}+\Delta\Omega_0^2.
\end{align}
For times outside the interaction region this always holds, since $\chi=0$. However, Eq. \eqref{ineqk} could fail for intermediate times if the strength of the interaction, regulated by the constant $\lambda$, is sufficiently high and $\Delta\Omega_0$ is also small. In that case the kernel-free part of the equation behaves locally as an overdamped oscillator, whose solutions in fact do not oscillate at all. This phenomenon in the time-independent vanishing-kernel case was already considered in \cite{Pipo}.

For the particular initial state $\hat{\rho}_\text{D} = \ket{0}\!\bra{0}$, i.e., if the detector is initially in its ground state defined by \mbox{$\hat{a}\ket{0} = 0$}, we find
\begin{align}
    \braket{e^{i\mu\hat{Q}_0(\tau)}}_{\text{D}} 
    &= \bra{0}e^{i\mu p(\tau)(u(\tau)\hat{a}+u^*(\tau)\hat{a}^\dagger)}\ket{0}
    \nonumber\\
    & = e^{-\mu^2 \mathfrak{H}(\tau)/2},
    \label{eq:hfrak}
    \end{align}
    where we have used Zassenhaus' formula \cite{Cohen-Tannoudji:101367} in the second equality, the fact that $a$ and $a^\dag$ satisfy the usual commutation relations for annihilation and creation operators, and we have defined 
\begin{align}
    \mathfrak{H}(\tau)= |p(\tau)u(\tau)|^2.
    \end{align}
We have chosen the ground state for convenience, but any other Gaussian state of the detector will have a characteristic function with a Gaussian dependence on $\mu$ as in Eq.~\eqref{eq:hfrak}, although the function $\mathfrak{H}(\tau)$  will possibly have a  more complicated, yet still quadratic, dependence on  $u,u^*$ \cite{Curtright2014}.

This allows us to write the induced observables for this spacetime and switching function as 
\begin{align}
    \mathcal{E}_\tau[\xi(\hat{Q})] =& \frac{1}{\sqrt{2\pi}}\int_{\mathbb{R}}d\mu \tilde{\xi}(\mu) 
   e^{-
   \mu^2 \mathfrak{H}(\tau)/2}e^{  i\mu \hat\varphi_0(\tau)} .
\end{align}
This expression can be written in a more convenient way in terms of convolutions with normalized Gaussian functions $ \mathcal{N}_{\sigma^2}(q)=  e^{-
   q^2/(2\sigma^2)}/\sqrt{2\pi\sigma^2}
$.
We can now undo the Fourier transform of $\xi$  and use the convolution formula to obtain
\begin{align}
    \mathcal{E}_\tau[\xi(\hat{Q})] =  \mathcal{N}_{\mathfrak{H}(\tau)}*\xi \big(\hat\varphi_0(\tau)\big).
\end{align}
The conclusion is that the observable induced by a function $\xi$ of the pointer variable $\hat{Q}$ is a new function of the smeared field $\hat\varphi_0(\tau)$. This new function is just a Gaussian filter applied to the original function with a time-dependent width $\sqrt{\mathfrak{H}(\tau)}$.

Whenever the width of the filter tends to zero, the function $\mathcal{N}_\mathfrak{H}$ tends to a $\delta$-function in the sense of distributions. In this limit, the induce observable acquires the simple form
\begin{align}
    \mathcal{E}_\tau[\xi(\hat{Q})] = \xi(\hat\varphi_0(\tau)),
    \label{eq:follow}
\end{align}
that is, the pointer variable is completely correlated with the smeared field amplitude. We can understand this limit then as an equilibration limit. Remarkably the equilibration timescales dictating this approximation are not related at all to the state of the field. This approximation of course depends on the function $\xi$ and its Fourier decomposition $\tilde \xi$. More specifically, we are assuming that the the filter-width $\sqrt{\mathfrak{H}(\tau)}$ is  much smaller than the width of $\tilde \xi$.
For fixed $\xi$ it holds in the limit $\int_{-\infty}^\tau d s\Gamma(s)\to \infty$.
For example, for times larger than the interaction time $T_\text{int}$, i.e. the width of $\chi$, this quantity becomes
\begin{align}
   \int_{-\infty}^\tau d s\Gamma(s)=\alpha\lambda^2\Gamma_0 T_\text{int},
   \label{eq:pinfty}
\end{align}
where $\alpha$ is a constant of order 1.  We then see that the detector follows the field in the sense of \eqref{eq:follow} and all the information about the initial state of the detector is erased in the limit $T_\text{int}\gg(\lambda^2\Gamma_0)^{-1}$.
 
\section{Systems with no memory kernel}\label{sec:nomem}
So far, we have written the induced observables in terms of fundamental solutions of Eq.~\eqref{eqn: finalintegrodif}, which is an integro-differential equation of second order. This equation has an integral kernel $K$ accounting for memory effects in the dynamics of the pointer variable. This is expected since it describes the full dynamics of an open system, which is not Markovian in general \cite{RivasHuelga2012}.  Finding analytical solutions to such equations for an arbitrary memory kernel is a very difficult task, even in if the kernel does not depend on the state of the environment, i.e. the quantum field, as in our case. However, there are situations where the memory kernel vanishes exactly, or to a good approximation. This is the case, for instance, for a massless field in (3+1)-dimensional Minkowski spacetime, where the retarded propagator equals its singular part and $K(\tau,s) = 0$.  One can interpret the absence of memory effects in this case through the so-called Huygens principle \cite{Hadamard1923}: As the detector interacts with the field, the perturbations induced by the interaction on the field travel through light-like geodesics, and these cannot have any influence on the detector, which follows a time-like geodesic. Therefore, from the point of view of the detector, the field is a memoryless environment. In more general spacetimes one could expect that, with the exception of some extreme situations, the backflow of information due to gravitational bouncing is negligible and that the memory kernel $K$ can be neglected for all practical purposes.  This can happen independently of the strength of the interaction, which is the usual figure of merit behind the derivation of Markovian master equations in general open quantum systems \cite{RivasHuelga2012}.

When the memory kernel vanishes,  $u$ is a complex solution to the equation
\begin{equation}\label{lau}
    \ddot u+\omega^2(\tau)u=0,
\end{equation}
normalized so that $u(\tau)\dot u^*(\tau)-u^*(\tau)\dot u(\tau)=i$ at any time~$\tau$, as we have seen.

It remains to analyze the form of the field amplitude that is being measured, its localization, and properties.
Given   two solutions of Eq.~\eqref{lau}, one can construct the retarded solution $R(\tau,s)$  as
\begin{align}
    &R(\tau,s) =-i\theta(\tau - s)\frac{p(\tau)}{2p(s)}\big(u(s)u^*(\tau) - u(\tau)u^*(s)\big).
\end{align}
Then, the induced field amplitude measured by the detector is just given by
\begin{align}
 \hat{\varphi}_0(\tau)
   & \nonumber=-ip(\tau)u(\tau)\int^\tau_{-\infty}\!\! d s\frac{\lambda\chi(s)u^{*}(s)}{2p(s)}\hat{\phi}_0(s)\\
   &+ip(\tau)u^*(\tau)\int^\tau_{-\infty} \!\!ds\frac{\lambda\chi(s)u(s)}{2p(s)}\hat{\phi}_0(s).
\end{align}

Note that for $\tau$ within the interaction region,the field is smeared against the discontinuous Heaviside function. This may jeopardize the definition of the field amplitude since the (pull-back of the) field is in general a well-defined operator-valued distribution on smooth functions~\cite{ReedSimon1975} but not necessarily on discontinuous ones. This problem is not surprising, since it is already present in perturbative calculations. However, if the processing time is taken to be after the interaction has finished, 
\begin{align}\label{amplitude}
     \hat{\varphi}_0(\tau)
   & \nonumber=-ip u(\tau)\int_{\mathbb{R}} ds\frac{\lambda\chi(s)u^{*}(s)}{p(s)}\hat{\phi}_0(s)\\
   &+ip u^*(\tau)\int_{\mathbb{R}}  ds \frac{\lambda\chi(s)u(s)}{p(s)}\hat{\phi}_0(s),
\end{align}
where $p$ does not depend on $\tau$ anymore outside the integral.

It is apparent from Eq.~\eqref{amplitude} that the induced amplitude is well defined, given that all the integrands are smooth and   compactly supported  if the switching function $\chi$ is. We also learn that the spacetime localization of this amplitude is indeed in the interaction region, That is, the segment of the trajectory of detector while $\chi$ is nonvanishing. Note that it has been argued \cite{fewster2020quantum} that statements about the localization of operators in QFT have to refer to causally convex regions, so it is perhaps more appropriate to say that the localization of the induced observable is in the causal diamond associated with the segment of the trajectory in which the interaction happens.

\subsection{Adiabatic limit}

The general form of the solution $u$ is
\begin{equation}
    u(\tau)=\frac{1}{\sqrt{2W(\tau)}}e^{ -i\int_{-\infty}^\tau ds W(s) },
\end{equation}
where $W$ satisfies the Ricatti equation
\begin{equation}
    W^2=\omega^2+ \left( \frac{3}{4}\frac{\dot{W}^2}{W^2}-\frac{1}{2}\frac{\ddot{W}}{W}     \right).
    \label{eq:ricatti}
\end{equation}
For adiabatic switching functions, $\omega$ varies very slowly and we can approximate $W(\tau)$ by $\omega(\tau)$ in the lowest adiabatic order \cite{Andrews1998}.

For times $\tau$ after the interaction has already finished, the filter-width becomes time-independent and takes the value
\begin{align}
 \mathfrak{H}(\tau)=\frac{\alpha\lambda^2\Gamma_0 T_\text{int}}{\sqrt{2\Omega_0}},
\end{align}
where $\alpha$ is a constant of order 1 (see Eq.~\eqref{eq:pinfty}). 
On the other hand,
it is straightforward to see that in the adiabatic regime and for times $\tau$ at which the interaction has already stopped,
\begin{align}
     \hat{\varphi}_0(\tau)
   & =\frac{-\lambda}{\sqrt{\Omega_0}}\int_\mathbb{R}  \!ds e^{-\lambda^2\Gamma_0\int_s^\infty \chi(t)^2 d t} \frac{\chi(s)}{ \sqrt{\omega(s)}}
   \nonumber\\
   &\qquad\qquad\times\sin\bigg(\int_{s}^\tau \!\!\! dt \omega(t)\bigg)\hat{\phi}_0(s).
\end{align}

\section{Non-Gaussianity}\label{sec: ng}

The goal of this section is to discuss how non-Gaussianities in the initial state of the field propagate to measurements induced by the interaction  (after it has already taken place) with the detector in an initial Gaussian state.

In general, deviations from Gaussianity can be characterized through the cumulant generating function
\begin{equation}\label{cumfir}
    \mathcal{C}_{\hat{Q}}(\mu;\tau) = \log[M_{\hat{Q}}(\mu;\tau)] = \sum_{n = 1}^{\infty}\kappa_{n}\frac{(i\mu)^n}{n!},
\end{equation}
where $\kappa_{n}$ is the $n$th cumulant.
By definition, Gaussian distributions have no cumulants of order greater that $n = 2$. More specifically, if some $\kappa_n$ for $n\geq 3$ does not vanish, the distribution of the pointer variable   is non-Gaussian.
This is especially interesting if we prepare the detector in an initial Gaussian state $\hat{\rho}_{\text{D}}$. We will now show that if, for some $n\geq3$, $\kappa_n\neq0$  after the interaction, the pointer variable is effectively recording non-Gaussianities in the field's initial state.

So far in this work, we have studied the detector's properties and how the field affects its dynamics, with little or no mention of the field's Hilbert space. We have mentioned, however, that the detector becomes correlated with a smeared field amplitude $\hat{\varphi}_0(\tau)$. Smeared field operators are the building blocks of rigorous approaches to quantum field theory in curved spacetimes \cite{Haag1996,FewsterRejzner2020}. 
The field is defined as an operator-valued distribution
\begin{align}
    \hat{\Phi}(f)=\int_\mathcal{M} dV \hat\Phi(\mathsf{x})f(\mathsf{x}),
\end{align}
where $f(\mathsf{x})$ are, in principle,  test functions, i.e. smooth.  Nonetheless, smeared field operators can also be defined for non-smooth  functions if the particular field satisfies some conditions \cite{Brunetti2000}. In our particular case, the field amplitude with which the detector correlates   can be written as
\begin{align}
   \hat{\varphi}_0(\tau)= \hat{\Phi}(j_\tau),
\end{align}
where $j_\tau$ is given by
\begin{align}
    j_\tau(\mathsf{x})=\int_\mathbb{R} d s \delta(\mathsf{x}-\mathsf{z}(s))\frac{-\lambda R(\tau,s)\chi(s)}{\sqrt{|\mathfrak{g}(\mathsf{z}(s))|}},
\end{align}
after the interaction has taken place. 
This is a well-defined distribution on spacetime functions. Indeed, it can be shown that $j_\tau$ is compatible with $\hat\Phi$ as distributions \cite{fewster2020quantum} provided that $\chi$ is a test function.

Gaussianity for field states can be defined analogously as for finite-dimensional systems. The (cumulant) generating function of the field for $\hat{\rho}_\phi$ is defined  as
\begin{align}
   \mathcal{C}_{\phi}[\mu,f]=\log (M_{\phi}[\mu,f]),\qquad M_{\phi}[\mu,f]=\braket{e^{i\mu\hat{\Phi}(f)}}_\phi.
\end{align}
 A state of the quantum field is said to be Gaussian with zero mean (sometimes called quasifree) if 
\begin{align}
    M_{\phi}[f]=e^{- {\mu^2}\mathcal{W}_{\hat{\rho}_\phi}[f,f]/{2}}
\end{align}
where
\begin{align}
    \mathcal{W}_{\hat{\rho}_\phi}[f,g]=\int_\mathcal{M} \!\!dV \int_\mathcal{M} \!\!dV' f(\mathsf{x})g(\mathsf{x}') \braket{\hat{\Phi}(\mathsf{x})\hat{\Phi}(\mathsf{x}')}_{\hat\rho_\phi}
\end{align}
is the so-called Wightman distribution of the field in the state $\hat{\rho}_\phi$, or two-point function. Equivalently, the cumulant generating function of a Gaussian state is quadratic in $\mu$ for all $f$. Examples of Gaussian states are vacuum and thermal states \cite{Wald1994}.
Any state that is not of this form is non-Gaussian.

The question is how non-Gaussianities in the field can be transferred to an originally Gaussian state for the detector. 
Consider the generating function of the pointer variable:
\begin{equation}\label{generating}
    M_{\hat{Q}}(\mu;\tau) = e^{-\mu^2{\mathfrak{H}(\tau)}/{2}}\braket{e^{i\mu \hat{\varphi}_0(\tau)}}_\phi,
\end{equation}
for some function $\mathfrak{H}(\tau)$ as discussed in Section~\ref{sec: indu}.  
Taking logarithms, we see that
\begin{align}\label{cumuint}
       \mathcal{C}_{\hat{Q}}(\mu)
     &  =
-\mu^2\frac{\mathfrak{H}(\tau)}{2}+\mathcal{C}_{\phi}[\mu,j_\tau].
\end{align}
It is clear that the cumulants of order $n\geq3$ coincide with the cumulants of the field exactly, since the detector's initial state only contributes to the $n=2$ order. Further, we observe that, given that $j_\tau$ is proportional to $\lambda$, the cumulant of order $n$ is of order $\mathcal{O}(\lambda^n)$, where $\lambda$ is the coupling constant. Therefore, the cumulant of order $n+1$ vanishes at the $n$th order in perturbation theory for $\lambda$. This implies, for example,  that calculations in second order in $\lambda$ (commonly the first nontrivial order) cannot distinguish Gaussian states of the field.

Let us now proceed with an example. A simple, yet ubiquitous, case of non-Gaussian states are one-particle states. 

Although the definition of one-particle states seems obvious in Minkowski spacetime, their definition in general backgrounds can be more involved. The reason is that these states require  a definition of  a set of annihilation operators $\hat a(f)$ and a vacuum vacuum state $\ket{0}$ such that 
\begin{align}
    \hat{\Phi}(f)=\hat{a}(f)+\hat{a}^\dagger(f),\qquad \hat{a}(f)\ket{0}=0
\end{align}
for all test functions $f$
\cite{Wald1994}. These creation and annihilation operators fulfill the commutation relations
\begin{align}
[\hat{a}(f),\hat{a}^\dagger(h)]=\mathcal{W}_{\ket{0}}[f,h]\openone, \qquad
    [\hat{a}(f),\hat{a}(h)]=0. 
\end{align}
Once we have chosen a vacuum state,  one-particle states are states of the form
\begin{align}
    \ket{g}=\hat{\Phi}(g)\ket{0},
\end{align}
where $g$ is normalized in the sense that $\mathcal{W}_{\ket{0}}[g,g]=1$.
The characteristic function of this state
 is easy to calculate taking into account the commutation relations of the creation and annihilation operators:
\begin{align}
    M_{g}[\mu, f]&=\braket{0|\hat{\Phi}(g)e^{i\mu\hat{\Phi}(f)}\hat{\Phi}(g)|0}
    \nonumber\\ &=\left(1-\mu^2\big|\mathcal{W}_{\ket{0}}[f,g]\big|^2\right)e^{- {\mu^2}\mathcal{W}_{\ket{0}}[f,f]/{2}}.
\end{align}
Therefore, its logarithm, i.e. the cumulant generating function of this one-particle state,
admits the following formal expansion in powers of $i\mu$:
\begin{align}\label{expfield}
   \mathcal{C}_{g}[\mu, f]&=(i\mu)^2\left(\frac12\mathcal{W}_{\ket{0}}[f,f]-\big|\mathcal{W}_{\ket{0}}[f,g]\big|^2\right)\nonumber\\
    &+\sum^\infty_{n=2}\frac{(-1)^n}{n}\big|\mathcal{W}_{\ket{0}}[f,g]\big|^{2n}(i\mu)^{2n}.
\end{align}
We see that the logarithm of the generating function has a power expansion involving infinitely many powers of $\mu$, so one-particle states are clearly non-Gaussian.

The cumulants for the detector interacting with a one-particle state of the field will be, comparing the power series expansions in \eqref{cumfir} and \eqref{expfield},
\begin{align}
   \kappa_2&=\mathfrak{H}(\tau)+\mathcal{W}_{\ket{0}}[j_\tau,j_\tau]-2\big|\mathcal{W}_{\ket{0}}[j_\tau,g]\big|^2, 
   \nonumber\\
\kappa_{2k}&=(-1)^k(k-1)!\big|\mathcal{W}_{\ket{0}}[j_\tau,g]\big|^{2k},
   \nonumber\\
   \kappa_{2k-1}&=0,
\end{align}
with $k\geq2$.
We see that all cumulants of the detector can be calculated and can be written in terms of the field's vacuum correlations smeared by $j_\tau$, which only depends on parameters characterizing the interaction and the classical equations of motion of the field. We can give the explicit expression of these   correlations 
\begin{align}
    \mathcal{W}_{\ket{0}}[j_\tau,g]= -\lambda \int_{\mathbb{R}} d s R(\tau,s)\chi(s)\mathcal{W}_{\ket{0}}[\mathsf{z}(s),g],
\end{align}
where
 $   \mathcal{W}_{\ket{0}}[\mathsf{x},g]=\braket{0|\hat{\Phi}(\mathsf{x})\hat{\Phi}(g)|0}
$.
In the case of a memoryless interaction, we can substitute further
\begin{align}\label{amplitude2}
     \mathcal{W}_{\ket{0}}[j_\tau,g]
   & \nonumber=-ip u(\tau)\int_{\mathbb{R}}\frac{\lambda\chi(s)u^{*}(s)}{p(s)}\mathcal{W}_{\ket{0}}[z(s),g]\\
   &+ip u^*(\tau)\int_{\mathbb{R}}\frac{\lambda\chi(s)u(s)}{p(s)}\mathcal{W}_{\ket{0}}[z(s),g],
\end{align}
from which we can distill some qualitative behavior.  These quantities are indeed smooth integrals of the positive energy solution $\mathcal{W}_{\ket{0}}[\mathsf{x},g]$, i.e. the Newton-Wigner wavefunction of the particle \cite{Fulling_1989}, along the trajectory. The integral is modulated by a smooth window $\chi/p$ regulating the time localization, and an oscillating function $u$, which probes the frequency content.

We should mention that the one-particle states we have analyzed to illustrate the effect of non-Gaussianities have a rather awkward behavior. They are non localizable, in the sense that they cannot be locally measured or generated \cite{BuchholzFredenhagen1982}. They are usually useful in the description of processes in which some nonlocal approximation has been made, e.g. long-time and perturbative approximations in particle physics or quantum optics. We cannot discard, however, that the type of effects that we are discussing in this paper, namely propagation of non-Gaussianities, lay outside the regime of validity of these states.
More reasonable states exhibiting non-Gaussianities can be locally generated, namely the vacuum of a field after undergoing any perturbation that is not quadratic in the fields, such as the one in the light-matter interaction. These non-Gaussianities can be generated, e.g., by a non-linear crystal as it is routinely done in quantum optical settings \cite{Boyd2019}. The description of such systems can be complex from a locality-friendly perspective, but our framework would be well suited to study them.

\section{Conclusions}\label{Sec: Conclusion}
In this work we have presented a  framework for discussing aspects of measurement and detection of quantum fields in the non-perturbative regime.  We have focused on building a model that, while amenable for non-perturbative calculations, is local, general covariant and causal. Therefore it applies for general trajectories in general globally hyperbolic spacetimes. 
We have focused our attention on finding an equation describing the dynamics of a pointer variable of a detector that couples to the field through a time-like trajectory, and characterized its singularities in the language of distribution theory.  We have renormalized the equation describing the dynamics in a covariant way, namely through the Epstein-Glasser renormalization procedure. 
This has allowed us to obtain a well-defined equation that depends on two phenomenological parameters (in (3+1)-dimensional spacetimes), namely a dispersion coefficient   and a Lamb shift. The renormalized equation is a linear integro-differential equation, which admits a formal solution in terms of independent  solutions to the homogeneous equation and a particular solution (the propagator acting on the source).   
 
 Given the formal solutions of the pointer variable's dynamics, we  have discussed how measurements of observables in the pointer variable induce measurements in the field, and we have characterized these observables for all functions of the pointer variable. Remarkably,   for initial Gaussian states of the detector, the observables induced by functions of the pointer variable can be calculated as a Gaussian filter applied to these functions, whose width depends on the  solution of the homogeneous integro-differential equation. Moreover, the functions of the pointer variable induce functions of a distinguished smeared field amplitude.  This amplitude is localized in the interaction region and the smearing is modulated by the detector's propagator. 
 
Moreover, we have also linked memory effects in the dynamics of the pointer variable with Huygens' principle, and we have seen that for situations in which the field propagates only through light-like geodesics, the (renormalized) evolution is memoryless. We then have considered the case of memoryless evolution, which can be formally solved in terms of solutions of a time-dependent frequency oscillator. This has allowed us to further factorize the homogeneous and particular solutions as a modulation and an oscillation, and to discuss the adiabatic approximation in this context.

Finally, we have analyzed an effect that cannot be described within perturbation theory at the lowest order, namely the recording of non-Gaussianities in the field through detector models. We have concluded that in our model, whenever the detector is prepared in a Gaussian state, the higher-order cumulants become nonvanishing if the initial state of the field is non-Gaussian. In fact, the cumulants of the detector evolve to be those of the field's state for a particular smeared field amplitude. We have analyzed the particular example of one-particle states, where we have calculated all the cumulants and their propagation into the detector, and interpreted them in terms of the vacuum's two-point function.

\acknowledgments

Financial support was provided by the Spanish Government Grants No. PID2023-149018NB-C44 and PID2023-148373NB-I00 (funded by MCIN/AEI/10.13039/501100011033  and by ``ERDF A way of making Europe''), the Q-CAYLE Project funded by the Regional Government of Castilla y Le\'on (Junta de Castilla y Le\'on) and by the Ministry of Science and Innovation (MCIN) through the European Union funds NextGenerationEU (PRTR C17.I1),  and the Natural Sciences and Engineering Research Council of Canada (NSERC).

\bibliography{biblio}

\end{document}